\documentclass[10pt,conference]{IEEEtran}
\usepackage{graphicx}
\usepackage{citesort}
\usepackage{epsfig}
\usepackage{fancybox}
\usepackage{color}
\usepackage[cmex10]{amsmath}
\usepackage{amsfonts}
\usepackage{cases}
\usepackage{url}
\usepackage{epstopdf}
\usepackage{subcaption}
\graphicspath{{./} {img/}}

\newtheorem{claim}{Claim}

\title{Signal Reconstruction in Linear Mixing Systems with Different Error Metrics}
\author{\authorblockN{Jin Tan and Dror Baron}
\authorblockA{Department of Electrical and Computer Engineering\\
North Carolina State University\\
Raleigh, NC 27695, USA \\
Email: \{jtan,dzbaron\}@ncsu.edu}
\thanks{This work was supported by the National Science Foundation, grant no. CCF-1217749, and by the U.S. Army Research Office, grant no. W911NF-04-D-0003. Portions appeared at the IEEE Statistical Signal Processing workshop (SSP), Aug. 2012~\cite{Tan2012SSP}, and additional manuscripts ~\cite{Tan2012signal,Tan2013}.}
}
\IEEEoverridecommandlockouts

\begin{document}
\maketitle
\newcommand{\xhat}{\widehat{\mathbf{x}}}
\newcommand{\xhati}{\widehat{x}_i}
\newcommand{\note}[1]{ {\bf *** #1 ***} }
\begin{abstract}
We consider the problem of reconstructing a signal from noisy measurements in linear mixing systems. The reconstruction performance is usually quantified by standard error metrics such as squared error, whereas we consider any additive error metric. Under the assumption that relaxed belief propagation (BP) can compute the posterior in the large system limit, we propose a simple, fast, and highly general algorithm that reconstructs the signal by minimizing the user-defined error metric. For two example metrics, we provide performance analysis and convincing numerical results. Finally, our algorithm can be adjusted to minimize the $\ell_\infty$ error, which is not additive. Interestingly, $\ell_{\infty}$ minimization only requires to apply a Wiener filter to the output of relaxed BP.
\end{abstract}
\section{Introduction}
\label{sec:Intro}
\subsection{Motivation}
\label{subsec:Motiv}
Linear mixing systems are popular models used in many settings such as compressed
sensing~\cite{CandesRUP,DonohoCS}, regression~\cite{Huber1973,OLeary1990}, and multiuser detection~\cite{GuoWang2008}. In this paper, we consider the following linear system~\cite{CandesRUP,DonohoCS,GuoWang2008},
\begin{eqnarray}
\label{eq:basicSystem}
\mathbf{w=\Phi x},
\end{eqnarray}
where the entries of the system input vector $\mathbf{x}\in\mathbb{R}^N$ are independent and identically distributed (i.i.d.), the random linear mixing matrix (or measurement matrix) $\mathbf{\Phi}\in\mathbb{R}^{M\times N}$ is sparse and known, and the measurement vector is $\mathbf{w}\in\mathbb{R}^M$. Because each component in $\mathbf{w}$ is a linear combination of the components in $\mathbf{x}$, we call the system \eqref{eq:basicSystem} a {\em linear mixing} system. 

The measurements $\mathbf{w}$ are passed through a bank of separable scalar channels characterized by conditional distributions,
\begin{eqnarray}
f_{\mathbf{Y|W}}(\mathbf{y|w})=\prod_{i=1}^M f_{Y|W}(y_i|w_i),
\label{eq:DisChannel}
\end{eqnarray}
where $\mathbf{y}$ is the channel output vector, and $(\cdot)_i$ denotes the $i$th element of the
corresponding vector. Note that the channels are general and are not restricted to Gaussian \cite{GuoWang2007,Rangan2010}. Our goal is to reconstruct the original system input $\mathbf{x}$ from the channel output $\mathbf{y}$ and the measurement matrix $\mathbf{\Phi}$. 

\subsection{Reconstruction quality}
\label{subsec:02}
The performance of the estimation process is often characterized by some additive error metric that quantifies the distance between the estimated signal and the original signal. For a signal $\mathbf{x}$ and its estimate $\widehat{\mathbf{x}}$, both of length $N$, the error between them is the summation over the component-wise errors,
\begin{eqnarray}
D(\mathbf{\widehat{x},x})=\sum_{j=1}^N d(\widehat{x}_j, x_j).
\label{eq:distDef}
\end{eqnarray}

Squared error, i.e., $d(\widehat{x}_j,x_j)=|\widehat{x}_j-x_j|^2$ in~\eqref{eq:distDef}, is one of the most popular error metrics in various problems, due to many of its mathematical advantages. For example, minimum mean squared error (MMSE) estimation provides both variance and bias information about an estimator~\cite{Grenander1957}, and in the Gaussian case it is linear and thus often easy to implement~\cite{Levy2008}. However, there are applications where MMSE estimation is inappropriate, for example it is sensitive to outliers~\cite{Cover91,Webb2002}. Therefore, alternative error metrics, such as mean absolute error (median), mean cubic error, or Hamming distance, are used instead. Considering the significance of various types of error metrics other than squared error, {\em a general estimation algorithm that can minimize any desired error metric is of interest}.

The error metric function~\eqref{eq:distDef} has an additive structure, and this particular structure makes it convenient to propose a scalar-estimation based algorithm. However, if we want to prevent any significant absolute errors, i.e., to minimize the $\ell_\infty$ norm error,
\begin{equation}
\|\mathbf{\widehat{x}-x}\|_\infty=\max_{j\in\{1,\ldots,N\}}|\widehat{x}_j-x_j|,\label{eq:ell_infty}
\end{equation}
where the error is not additive, then {\em is it possible to utilize the additive error metric~\eqref{eq:distDef} to approximate a minimum mean~$\ell_\infty$ error estimator}?

\subsection{Related work}
\label{subsec:Related}
Squared error is most commonly used as the error metric in estimation problems given by~\eqref{eq:basicSystem} and~\eqref{eq:DisChannel}. Mean-square optimal analysis and algorithms were introduced in~\cite{GuoVerdu2005,Guo2006,GuoBaronShamai2009,RFG2009,CSBP2010} to estimate a signal from measurements corrupted by Gaussian noise; in~\cite{GuoWang2007,Rangan2010,RanganGAMP2010}, further discussions were made about the circumstances where the output channel is arbitrary, while, again, the MMSE estimator was put forth. Another line of work, based on a greedy algorithm called {\em orthogonal matching pursuit}, was presented in~\cite{TroppOMP,Cosamp08} where the mean squared error decreases over iterations. Absolute error is also under intense study in signal estimation. For example, an efficient sparse recovery scheme that minimizes the absolute error was provided in~\cite{indyk2008near,berinde2008practical}; in~\cite{CDDNOA}, a fundamental analysis was offered on the minimal number of measurements required while keeping the estimation error within a certain range, and absolute error was one of the metrics concerned. Support recovery error is another metric of importance, for example it relates to properties of the measurement matrices~\cite{Wang2010}.
The authors of~\cite{Tulino2011,Wainwright2009,Wang2010} discussed the support error rate when recovering a sparse signal from its noisy measurements; support-related performance metrics were applied in the derivations of theoretical limits on the sampling rate for signal recovery~\cite{Akcakaya2010,Reeves2011sampling}.

There have also been a number of studies on general properties of $\ell_\infty$ error related solutions. An overdetermined linear system $\mathbf{y=\Phi x}$, where $\mathbf{\Phi}\in\mathbb{R}^{N\times M}$ and $N>M$, was considered by Cadzow~\cite{cadzow2002}, and the properties of the minimum $\ell_\infty$ error solutions to this system was explored. In Clark~\cite{clark1961}, the author developed a way to calculate the distribution of the greatest element in a finite set of random variables. And in Indyk~\cite{indyk2001}, an algorithm was introduced to find the nearest neighbor of a point while $\ell_\infty$ norm distance was considered. Finally, Lounici~\cite{Lounici2008} studied the $\ell_\infty$ error convergence rate for Lasso and Dantzig estimators.
\subsection{Contributions}
\label{sec:contrib}

In this paper, we review our recent work on signal estimation with arbitrary additive error metrics~\cite{Tan2012SSP,Tan2012signal}, and discuss its extension to minimizing the $\ell_\infty$ norm error. Our contributions are: ({\em i}) we suggest a Bayesian estimation algorithm that minimizes an arbitrary additive error metric; ({\em ii}) we prove that the algorithm is optimal, and study the fundamental information-theoretic performance limit of estimation for a given metric; and ({\em iii}) we derive the performance limits for
minimum mean absolute error and minimum mean support error.

The proposed metric-optimal algorithm is based on the assumption that the relaxed belief propagation (BP) method~\cite{Rangan2010} converges to a set of degraded scalar Gaussian channels~\cite{GuoVerdu2005,GuoBaronShamai2009,RFG2009,GuoWang2007}.
The relaxed BP method is well-known to be optimal for the squared error, while we further show that the relaxed BP method can do more -- because the sufficient statistics are given, other additive error metrics can also be minimized with one extra step, which is simple and fast.
This is convenient for users who desire to recover the original signal with a non-standard additive error metric.

Another contribution that is included in this paper is an application of the metric-optimal algorithm to the~$\ell_\infty$ norm error, where the input signal is {\em sparse Gaussian} distributed, meaning that with probability $s$ the vector entries are standard Gaussian, and with probability $(1-s)$ they are zero.
We find~\cite{Tan2013} that, in the large system limit, the minimum mean~$\ell_\infty$ estimator is a linear function of the output of the relaxed belief propagation (BP) method. Moreover, based on the metric-optimal algorithm, we propose a heuristic that achieves a lower $\ell_\infty$ error than the linear function does when the signal dimension is finite. 

The remainder of the paper is arranged as follows. We review the relaxed BP method in Section~\ref{sec:Review}. Then we describe our proposed algorithm, discuss its performance, and provide several example error metrics in Section~\ref{sec:EstSys}. We further discuss the possibility of extending the proposed algorithm for $\ell_\infty$ norm error in Section~\ref{sec:ell_infty}. Simulation results appear in Section~\ref{sec:Sim}.

\section{Review of Relaxed Belief Propagation}
\label{sec:Review}

\begin{figure}[t]
\centering
\includegraphics[width=5cm]{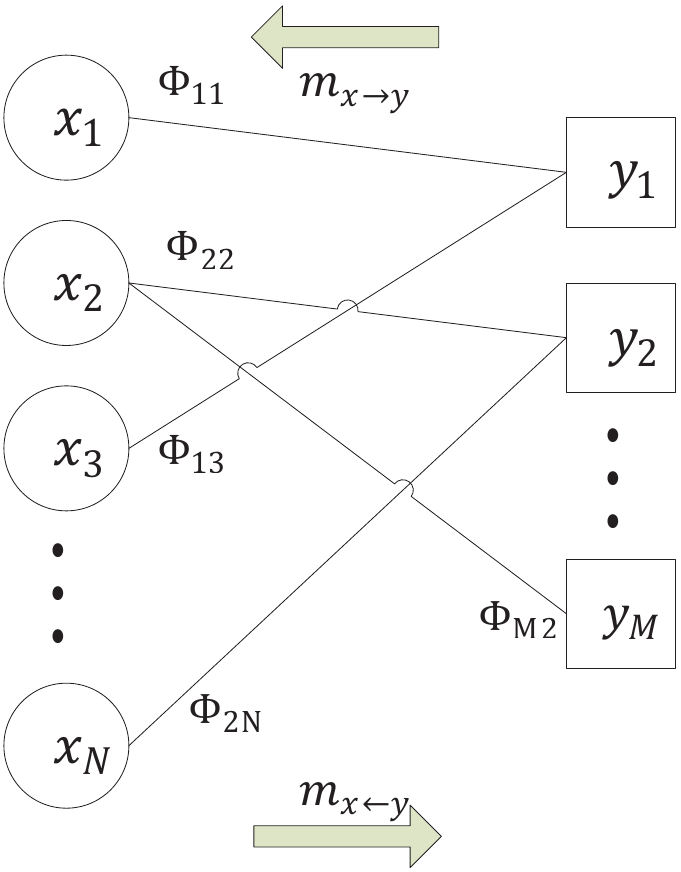}
\caption{
{\small\sl
Tanner graph for relaxed belief propagation.
}
\label{fig:RelaxBP}
}
\end{figure}
Belief Propagation (BP)~\cite{Bishop2006} is an iterative method used to compute the marginal probabilities of a Bayesian network. Consider the bipartite graph, called a \textit{Tanner} or \textit{factor} graph, shown in Figure \ref{fig:RelaxBP}, where circles represent random variables (called \textit{variable nodes}), and related variables are connected through functions (represented by squares, called \textit{factor nodes} or \textit{function nodes}) that indicate their Bayesian dependence. In standard BP, there are two types of messages passed through the nodes: messages from variable nodes to factor nodes, $m_{x\rightarrow y}$, and messages from factor nodes to variable nodes, $m_{y\rightarrow x}$. If we denote the set of function nodes connected to the variable $x$ by $N(x)$, the set of variable nodes connected to the function $y$ by $N(y)$, and the factor function at node $y$ by $\Psi_y$, then the two types of messages are defined as follows~\cite{Bishop2006}:
\begin{eqnarray}
m_{x\rightarrow y}&=&\prod_{k\in N(x)\setminus y}m_{k\rightarrow x},\nonumber\\
m_{y\rightarrow x}&=&\sum_{\ell\in N(y)\setminus x}\Psi_y m_{\ell\rightarrow y}\nonumber,
\end{eqnarray}
where $A\setminus B$ denotes the set whose elements are in $A$ but not in $B$.

Inspired by the basic BP idea described above, the authors of~\cite{Guo2006,GuoWang2007} developed iterative algorithms for estimation problems in linear mixing systems. In the Tanner graph, an input vector $x=[x_1,x_2,...,x_N]$ is associated with the variable nodes (input nodes), and the output vector $y=[y_1,y_2,...,y_M]$ is associated with the function nodes (output nodes). If~$\Phi_{ij}\neq0$, then nodes~$x_j$ and~$y_i$ are connected by an edge~$(i,j)$, where the set of such edges~$E$ is defined as~$E=\{(i,j): \Phi_{ij}\neq0\}$.

In standard BP methods~\cite{Caire2004,Montanari2006,GuoWang2008}, the distribution functions of $x_j$ and $w_i$ as well as the channel distribution function $f_{Y|W}(y_i|w_i)$ were set to be the messages passed along the graph, but it is difficult to compute those distributions, making the standard BP method computationally expensive. In~\cite{Guo2006}, a simplified algorithm, called \textit{relaxed BP}, was suggested. In this algorithm, means and variances replace the distribution functions themselves and serve as the messages passed through nodes in the Tanner graph, greatly reducing the computational complexity. In~\cite{GuoWang2007,Rangan2010,RanganGAMP2010,Rangan2011}, this method was extended to a more general case where the channel is not necessarily Gaussian.

In Rangan~\cite{Rangan2010}, the relaxed BP algorithm generates two sequences, $q_j(t)$ and $\mu_j(t)$, where $t\in\mathbb{Z}^+$ denotes the iteration number. Under the assumptions that the signal dimension $N\rightarrow\infty$ and the iteration number $t\rightarrow\infty$, while the ratio $M/N$ is fixed, the sequences $q_j(t)$ and $\mu_j(t)$ converge to sufficient statistics for the linear mixing channel observation $\mathbf{y}~\eqref{eq:DisChannel}$. More specifically, the conditional distribution $f(x_j|q_j(t),\mu_j(t))$ converges to the conditional distribution $f(x_j|\mathbf{y})$, where $q_j(t)$ can be regarded as a Gaussian-noise-corrupted version of $x_j$, and $\mu_j(t)$ is the noise variance,
\begin{equation}
\label{eq:1st_scalarGchannel}
q_j(t)=x_j+v_j(t),
\end{equation}
where $v_j(t)\sim\mathcal{N}(0,\mu_j(t))$
for $j=1,2,\ldots,N$. It has been shown~\cite{Rangan2010} that $\mu_j(t)$ converges to a fixed point that satisfies Tanaka's equation, which is analyzed in detail in~\cite{Tanaka2002,GuoVerdu2005,Montanari2006,GuoBaronShamai2009,RFG2009,DMM2009,RanganGAMP2010}. We define the limits of the sequences,
\begin{eqnarray}
\lim_{t\rightarrow\infty}q_j(t)&=&q_j,\nonumber\\
\lim_{t\rightarrow\infty}v_j(t)&=&v_j,\nonumber\\
\lim_{t\rightarrow\infty}\mu_j(t)&=&\mu,\nonumber
\end{eqnarray}
for $j=1,2,\ldots,N$. Note that all the scalar Gaussian channels~\eqref{eq:1st_scalarGchannel} converge to the same noise variance $\mu$. We now simplify equation~\eqref{eq:1st_scalarGchannel} as,
\begin{equation}
q_j=x_j+v_j,
\label{eq:scalarGchannel}
\end{equation}
where $v_j\sim\mathcal{N}(0,\mu)$
for $j=1,2,\ldots,N$.

\section{Estimation Algorithm for Additive Errors}
\label{sec:EstSys}
\subsection{Algorithm}
\begin{figure}[t]
\centering
\includegraphics[width=9cm]{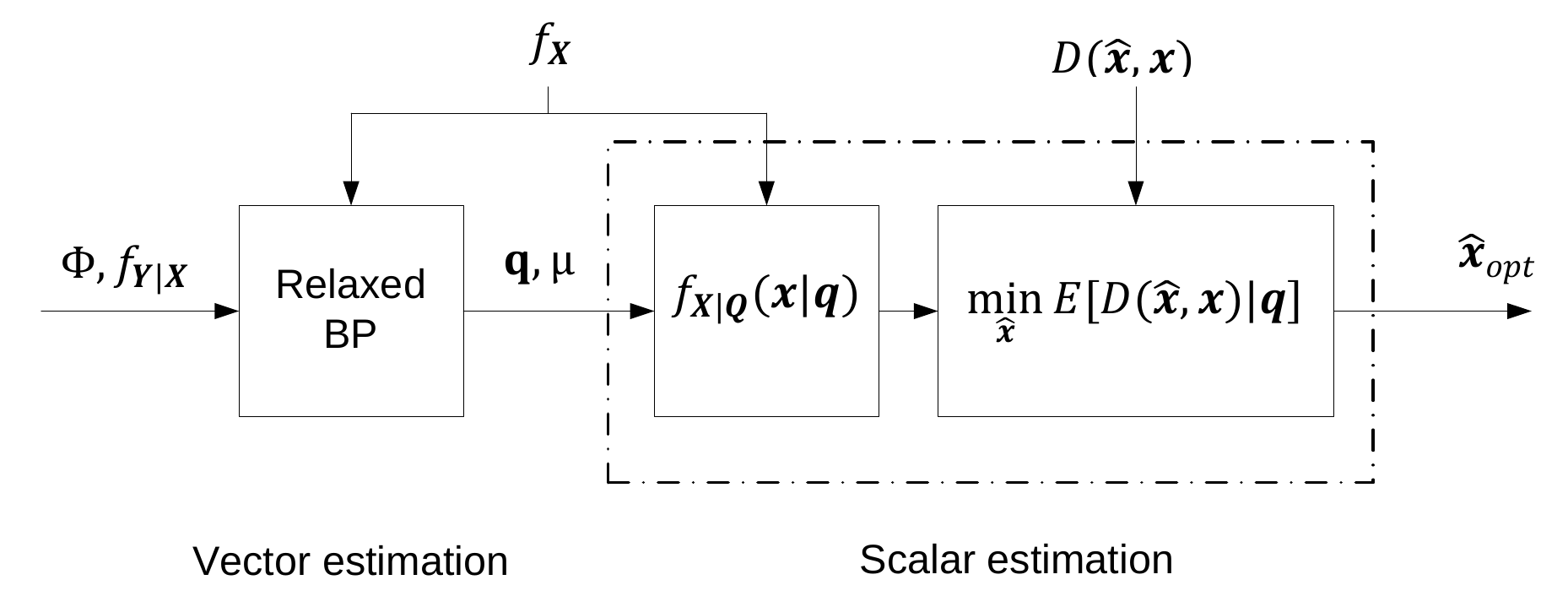}
\caption{
{\small\sl
The structure of the metric-optimal estimation algorithm.
}
\label{fig:EstSys}
}
\vspace*{-5mm}
\end{figure}
In this paper, we utilize the outputs of the relaxed BP method by Rangan~\cite{Rangan2010}, in particular the software package ``GAMP"~\cite{Rangan:web:GAMP}.
Figure \ref{fig:EstSys} illustrates the structure of our 
{\em metric-optimal estimation algorithm} (dashed box). The inputs of the algorithm are: ({\em i}) a distribution function $f_{\mathbf{X}}(\mathbf{x})$, the prior of the original input $\mathbf{x}$; {(\em ii)} a vector $\mathbf{q}=[q_1,q_2,...,q_N]$, the outputs of the scalar Gaussian channels computed by relaxed BP~\cite{Rangan2010}; ({\em iii}) a scalar $\mu$, the variance of the Gaussian noise in \eqref{eq:scalarGchannel}; and ({\em iv}) an additive error metric function $D(
\mathbf{\widehat{x},x})$~\eqref{eq:distDef} specified by the user.

Because the scalar channels have additive Gaussian noise, and the variances of the noise are all $\mu$, we can compute the conditional probability density function $f_{\mathbf{X|Q}}(\mathbf{x|q})$ from Bayes' rule:
\begin{eqnarray}
f_{\mathbf{X|Q}}(\mathbf{x|q})&=&\frac{f_{\mathbf{Q|X}}(\mathbf{q|x})f_\mathbf{X}(\mathbf{x})}{f_\mathbf{Q}(\mathbf{q})}\nonumber\\
&=&\frac{f_{\mathbf{Q|X}}(\mathbf{q|x})f_\mathbf{X}(\mathbf{x})}{\int f_{\mathbf{Q|X}}(\mathbf{q|x})f_\mathbf{X}(\mathbf{x}) d\mathbf{x}},
\label{eq:BayesRule}
\end{eqnarray}
where
\begin{equation}
f_{\mathbf{Q|X}}(\mathbf{q|x})=\frac{1}{\sqrt{(2\pi\mu)^N}}\exp\left(-\frac{\|\mathbf{q-x}\|_2^2}{2\mu}\right).\nonumber
\end{equation}
Given an error metric $D(\mathbf{\widehat{x},x})$, the optimal estimand $\widehat{\mathbf{x}}_{\text{opt}}$ is generated by minimizing the conditional expectation of the error metric $E[D(\mathbf{\widehat{x},x)|q}]$, which is easy to compute using $f_{\mathbf{X|Q}}(\mathbf{x|q})$:
\begin{equation}
E[D(\mathbf{\widehat{x},x)|q}]=\int D(\mathbf{\widehat{x},x})f_{\mathbf{X|Q}}(\mathbf{x|q})d\mathbf{x}.\nonumber
\label{eq:conditionalExp}
\end{equation}
Then,
\begin{eqnarray}
\widehat{\mathbf{x}}_{\text{opt}}&=&\arg\min_{\mathbf{\widehat{x}}} E[D(\mathbf{\widehat{x},x)|q}]\nonumber\\
&=&\arg\min_{\mathbf{\widehat{x}}} \int D(\mathbf{\widehat{x},x})f_{\mathbf{X|Q}}(\mathbf{x|q})d\mathbf{x}.
\label{eq:mainAlg}
\end{eqnarray}

The conditional probability $f_{\mathbf{X|Q}}(\mathbf{x|q})$ is separable, because the parallel scalar Gaussian channels~\eqref{eq:scalarGchannel} are separable and $f_{\mathbf{X}}(\mathbf{x})$ is i.i.d. Moreover, the error metric function $D(\mathbf{\widehat{x},x})$~\eqref{eq:distDef} is also separable. Therefore, the problem reduces to scalar estimation~\cite{Levy2008},
\begin{eqnarray}
\widehat{x}_{\text{opt},j}&=&\arg\min_{\widehat{x}_j} E[d(\widehat{x}_j,x_j)|q_j]\nonumber\\
&=&\arg\min_{\widehat{x}_j} \int d(\widehat{x}_j,x_j)f_{x|q}(x_j|q_j)dx_j,
\label{eq:scalarEst}
\end{eqnarray}
for $j=1,2,\ldots,N$.
Equation~\eqref{eq:scalarEst} minimizes a single-variable function.
We have shown in detail~\cite{Tan2012SSP,Tan2012signal} how to perform this minimization in two example error metrics, and summarize the results in Section~\ref{subsec:Eg}.
\newcounter{mytempeqncnt}
\begin{figure*}[!t]
\normalsize
\setcounter{mytempeqncnt}{\value{equation}}
\begin{eqnarray}
\text{MMUE}(f_\mathbf{X},\mu)&=&\int_{R(\mathbf{Q})}\int_{R(\mathbf{X})}D(\widehat{\mathbf{x}}_{\text{opt}},\mathbf{x})\left(\frac{1}{\sqrt{(2\pi\mu)^{N}}}\exp\left({-\frac{\|\mathbf{q}-\mathbf{x}\|^2}{2\mu}}\right)\right)f_\mathbf{X}(\mathbf{x})d\mathbf{x}d\mathbf{q}.
\label{eq:thmbound}\\
\text{MMAE}(f_{\bf{X}},\mu)
&=&N\int_{-\infty}^{+\infty}\left(\int_{-\infty}^{\widehat{x}_{j,\text{MMAE}}} (-x_j)f_{X_j|Q_j}(x_j|q_j)dx_j
+\int_{\widehat{x}_{j,\text{MMAE}}}^{+\infty} x_j f_{X_j|Q_j}(x_j|q_j)dx_j\right)f_{Q_j}(q_j)dq_j.
\label{eq:MMAEBound}
\end{eqnarray}
\hrulefill
\vspace*{4pt}
\end{figure*}
\subsection{Theoretical results}
\label{subsec:TheoResult}

Having described the algorithm, we now discuss its theoretical properties based on the replica method. Because the replica method has not been rigorously justified, our results are given as claims. More detailed discussions of the claims can be found in Tan et al.~\cite{Tan2012signal}.

\begin{claim}
Given the system model~\eqref{eq:basicSystem}, \eqref{eq:DisChannel} and an error metric $D(\mathbf{\widehat{x},x})$ of additive form~\eqref{eq:distDef}, as the signal dimension $N\rightarrow\infty$ and the measurement ratio $M/N$ is fixed, the optimal estimand of the input signal is given by
\begin{equation}
\mathbf{\widehat{x}}_{\text{opt}}=\arg\min_{\mathbf{\widehat{x}}} E\left[D(\mathbf{\widehat{x},x)|q}\right],\nonumber
\label{eq:thmEq}
\end{equation}
where the vector entries $\mathbf{q}=(q_1, q_2, \ldots, q_N)$ are the outputs of the scalar Gaussian channels \eqref{eq:scalarGchannel}.
\label{thm:myThm}
\end{claim}

As we mentioned in Section~\ref{sec:Review}, the probability density function $f_{X_j|\mathbf{Y}}(x_j|\mathbf{y})$ is statistically equivalent to $f_{X_j|Q_j}(x_j|q_j)$ in the large system limit. Under this assumption,
once we know the value of $\mu$, estimating each $x_j$ from all channel outputs $\mathbf{y}=(y_1,y_2,...,y_M)$ is equivalent to estimating $x_j$ from the corresponding scalar channel output $q_j$. The relaxed BP method~\cite{Rangan2010} calculates the sufficient statistics $q_j$ and $\mu$. Therefore, an estimator based on minimizing the conditional expectation of the error metric, $E\left(D(\mathbf{\widehat{x},x)|q}\right)$, gives an asymptotically optimal result.

\begin{claim}
With the optimal estimand $\mathbf{\widehat{x}}_{\text{opt}}$ determined by \eqref{eq:mainAlg}, the minimum mean user-defined error (MMUE) is given by~\eqref{eq:thmbound}, where $R(\cdot)$ represents the range of a variable, and $\mu$ is the variance of the noise of the scalar Gaussian channel \eqref{eq:scalarGchannel}.
\label{thm:BoundExpression}
\end{claim}

\subsection{Examples}
\label{subsec:Eg}
We include examples that have been worked out in detail in Tan et al.~\cite{Tan2012SSP,Tan2012signal}.

\subsubsection{Absolute error}
\label{subsub:MAE}
Because the MMSE is the mean of the conditional distribution, the outliers in the set of data may corrupt the estimation, and in this case the minimum mean absolute error (MMAE) could be a good alternative. The minimum mean absolute error is given by~\eqref{eq:MMAEBound},
where $x_i$ (respectively, $q_i$) is the input (respectively, output) of the scalar Gaussian channel~\eqref{eq:scalarGchannel}, $\widehat{x}_{i,\text{MMAE}}$ is such that $\int_{\widehat{x}_{i,\text{MMAE}}}^{+\infty}f_{X_i|Q_i}(x_i|q_i)dx_i=\frac{1}{2}$, and $f_{X_i|Q_i}(x_i|q_i)$ is a function of $f_{X_i}(x_i)$ following~\eqref{eq:BayesRule}.

\subsubsection{Support recovery error}
\label{subsub:support}
In some applications, correctly estimating the locations where the data has non-zero values is almost as important as estimating the exact values of the data; it is a standard model selection error criterion~\cite{Wang2010}. The process of estimating the non-zero locations is called \textit{support recovery}. Support recovery error is defined as follows, and this metric function is discrete,
\begin{eqnarray}
d_{\text{support}}(\widehat{x}_j,x_j)=\text{xor} (x_j, \widehat{x_j}),\nonumber
\end{eqnarray}
where xor is the standard logical operator.

We have shown that the minimum mean support error (MMSuE) estimator  achieves
\begin{eqnarray}
\text{MMSuE}(f_{\mathbf{X}},\mu)&=&N(1-s)\cdot\text{erfc}\left(\sqrt{\frac{\tau}{2\mu}}\right)\nonumber\\
&+&Ns\cdot\text{erf}\left(\sqrt{\frac{\tau}{2(\sigma^2+\mu)}}\right),\label{eq:supportBound}
\end{eqnarray}
where
\begin{equation}
\tau=2\cdot\frac{\sigma^2+\mu}{\sigma^2/\mu}\ln\left(\frac{(1-s)\sqrt{\sigma^2/\mu+1}}{s}\right).\nonumber
\end{equation}

\section{The Minimum Mean $\ell_\infty$ Error Estimator}
\label{sec:ell_infty}

Having discussed the metric-optimal algorithm, which was designed for additive error metrics, we now discuss how to extend the algorithm to support $\ell_\infty$ error.

The $\ell_\infty$ error only considers the component with greatest absolute value~\eqref{eq:ell_infty}, and the minimum mean~$\ell_\infty$ error estimator is defined as
\begin{equation}
\bf{\widehat{x}}_\infty=\arg\min_{\bf{\widehat{x}}}E\left[\|\bf{\widehat{x}-x}\|_\infty|\bf{q}\right].
\label{eq:min_infty}
\end{equation}

For a scalar Gaussian channel \eqref{eq:scalarGchannel}, we know that the minimum mean squared error estimator is achieved by conditional expectation. If $x_j\sim\mathcal{N}(0,1)$, then it is the estimator $\frac{1}{1+\mu}\bf{q}$ that gives the minimum mean squared error, where $\mu$ is the variance of the noise in the scalar Gaussian channels~\eqref{eq:scalarGchannel}. This format $\frac{1}{1+\mu}\bf{q}$ is regarded as the {\em Wiener filter} in signal processing~\cite{Wiener1949}. Our ongoing work~\cite{Tan2013} indicates that, in the large system limit, the Wiener filter is also optimal for $\ell_\infty$ error when the input signal is {\em sparse} Gaussian. 

\begin{claim}
For a set of parallel scalar Gaussian channels~\eqref{eq:scalarGchannel}, where the input signal is sparse Gaussian distributed, the minimum mean $\ell_\infty$ error estimator~$\widehat{\mathbf{x}}_\infty\in\mathbb{R}^N$~\eqref{eq:min_infty} is linear in the channel observation vector~$\mathbf{q}$ as $N\rightarrow\infty$, i.e.,
\begin{equation}
\lim_{N\rightarrow\infty}\mathbf{\widehat{x}}_\infty=\frac{\mu_x}{\mu_x+\mu_v}\mathbf{q}.
\label{eq:c_q}
\end{equation}
\label{thm:min_infty}
\end{claim}

The detailed proof of Claim~\ref{thm:min_infty} is in preparation~\cite{Tan2013}.

Unfortunately, the Wiener filtering~\eqref{eq:c_q} is only optimal for $\ell_\infty$ error in the large system limit, and when the signal dimension is finite, our metric-optimal algorithm in conjunction with a specific error metric {\em outperforms} the Wiener filtering in terms of $\ell_\infty$ error. 

Let us see what this specific error metric is.
Recall that the definition of the $\ell_p$ norm error between $\mathbf{\widehat{x}}$ and $\mathbf{x}$ is
\begin{equation}
\|\widehat{\mathbf{x}}-\mathbf{x}\|_p=\left(\sum_{j \in \{1,\ldots,N\} } |\widehat{x}_j-x_j|^p\right)^{1/p},\nonumber
\end{equation}
and it can be shown that
\begin{equation}
\lim_{p\rightarrow +\infty}\|\widehat{\mathbf{x}}-\mathbf{x}\|_p= \|\widehat{\mathbf{x}}-\mathbf{x}\|_\infty.\nonumber
\label{eq:p_and_infty}
\end{equation}
It is therefore natural to turn to the $\ell_p$ norm error as an alternative.

The $\ell_p$ error is closely related to our definition of an additive error metric~\eqref{eq:distDef}. We define
\begin{equation}
D_p(\mathbf{\widehat{x},x})=\sum_{j=1}^N |\widehat{x}_j- x_j|^p=\|\mathbf{\widehat{x}-x}\|_p^p\label{eq:metricDef_lp},\nonumber
\end{equation}
and let $\widehat{\mathbf{x}}_p$ denote the estimand that minimizes the conditional expectation of $D_p(\mathbf{\widehat{x},x})$, i.e.,
\begin{eqnarray}
\widehat{\mathbf{x}}_p&=&\arg\min_{\widehat{\mathbf{x}}} E[D_p(\mathbf{\xhat,x})|\mathbf{q}]\nonumber\\
&=&\arg\min_{\widehat{\mathbf{x}}} E[\|\mathbf{\widehat{x}-x}\|_p^p|\mathbf{q}],
\label{eq:xhat_p}
\end{eqnarray}
and
\begin{equation}
\widehat{x}_{p,j}=\arg\min_{\widehat{x}_j} E[|\widehat{x}_j-x_j|^p|q_j],\nonumber
\label{eq:xhat_pi}
\end{equation}
for $j\in\{1,2,...,N\}$.

The numerical results in Section~\ref{sec:Sim} show that, for some values of $p$, the estimator~$\mathbf{\widehat{x}}_p$~\eqref{eq:xhat_p} achieves a lower~$\ell_\infty$ error than the Wiener filter~\eqref{eq:c_q}.

\section{Numerical results}
\label{sec:Sim}

To illustrate the performance of our algorithms, we provide some numerical results in this section, for both additive error metrics and the $\ell_\infty$ norm error. The Matlab implementation of our algorithm can be found at \url{http://people.engr.ncsu.edu/dzbaron/software/arb_metric/}. It automatically computes equations~\eqref{eq:BayesRule}-\eqref{eq:scalarEst} where the additive error metric~\eqref{eq:distDef} is given as the input of the algorithm.

\subsection{Additive error metrics}
\label{subsec:additive_Sim}
We test our estimation algorithm on two cases modeled by \eqref{eq:basicSystem} and \eqref{eq:DisChannel}: ({\em i}) sparse Gaussian input and Gaussian channel; ({\em ii}) sparse Weibull (defined below) input and Poisson channel. In both cases, the input's length $N$ is 10,000, and its sparsity rate is $s=3\%$, meaning that the entries of the input vector are non-zero with probability $3\%$, and zero otherwise. The matrix~$\mathbf{\Phi}$ we use is Bernoulli($0.5$) distributed, and is normalized to have unit-norm rows. In the first case, the non-zero input entries are $\mathcal{N}(0,1)$ distributed, and the Gaussian noise is $\mathcal{N}(0,3\cdot 10^{-4})$ distributed, i.e., the signal-to-noise ratio (SNR) is 20dB. In the second case, the non-zero input entries are Weibull distributed,
\begin{equation}
f(x_j;\lambda,k)=
\begin{cases}
\frac{k}{\lambda}\left(\frac{x_j}{\lambda}\right)^{k-1}e^{-(x_j/\lambda)^k}&x_j\ge 0\\
0&x_j<0
\end{cases},\nonumber
\end{equation}
where $\lambda=1$ and $k=0.5$. The Poisson channel is
\begin{equation}
f_{Y|W}(y_i|w_i)=\frac{(\alpha w_i)^{y_i}e^{-(\alpha w_i)}}{y_i!},\quad i\in\{1,2,\ldots,M\},\nonumber
\end{equation}
where the scaling factor of the input is $\alpha=100$.

In order to illustrate that our estimation algorithm is suitable for reasonable error metrics, we considered absolute error and two other non-standard metrics:
\begin{equation}
\label{eq:OtherMetric}
\text{Error}_p=\sum_{j=1}^N |\widehat{x}_j-x_j|^p,\nonumber
\end{equation}
where $p=0.5$ or $1.5$.

\begin{figure}[t]
\centering
\includegraphics[width=8cm]{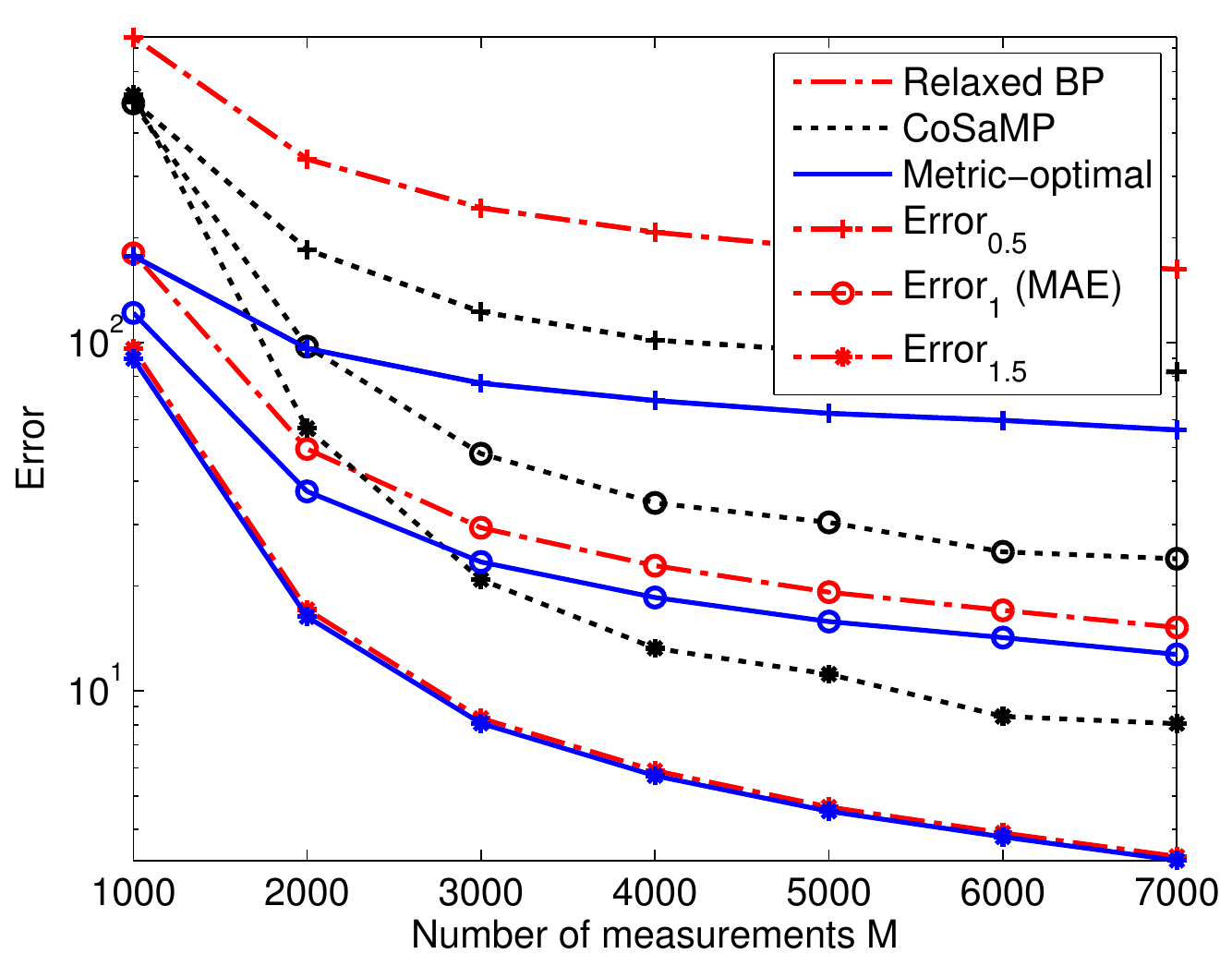}
\caption{
{\small\sl
Comparison of the metric-optimal estimation algorithm, relaxed BP, and CoSaMP. (Sparse Gaussian input and Gaussian channel; sparsity rate~$s=3\%$; input length N=10,000; SNR=20dB.)
}
\label{fig:PlotG}
}
\end{figure}

\begin{figure}[t]
\centering
\includegraphics[width=8.6cm]{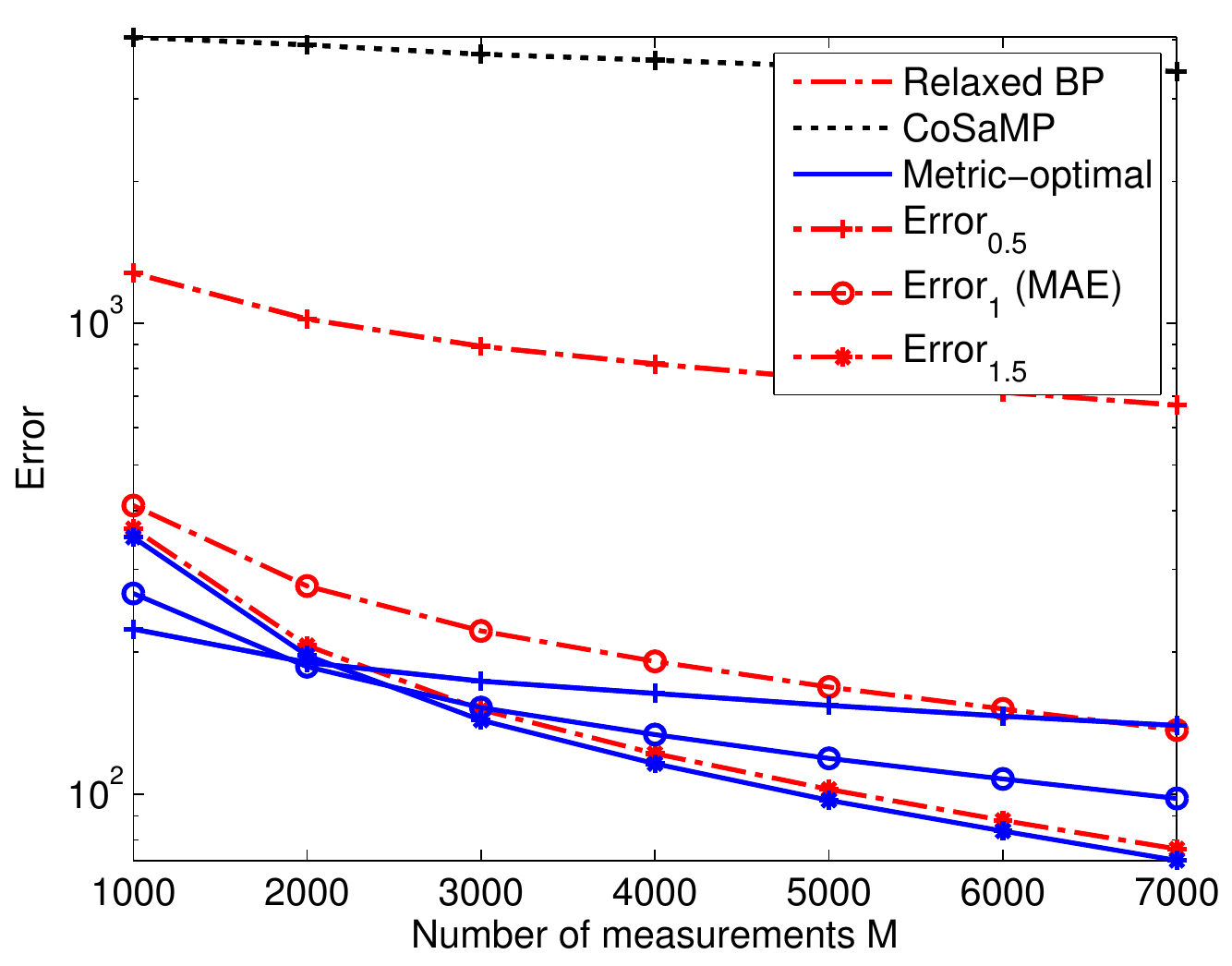}
\caption{
{\small\sl
Comparison of the metric-optimal estimation algorithm, relaxed BP, and CoSaMP. The ``MAE" and the ``$\text{Error}_{1.5}$" lines for ``CoSaMP" appear beyond the scope of the vertical axis. (Sparse Weibull input and Poisson channel; sparsity rate~$s=3\%$; input length N=10,000; input scaling factor $\alpha=100$.)
}
\label{fig:PlotW}
}
\end{figure}

We compare our algorithm with the relaxed BP~\cite{Rangan2010} and CoSaMP~\cite{Cosamp08} algorithms. In Figure \ref{fig:PlotG} and Figure \ref{fig:PlotW}, lines marked with ``metric-optimal" present the errors of our estimation algorithm, and lines marked with ``Relaxed BP" (respectively, ``CoSaMP") show the errors of the relaxed BP (respectively, CoSaMP) algorithm. Each point in the figure is an average of 100 experiments with the same parameters. Because the Poisson channel is not an additive noise channel and is not suitable for CoSaMP, the ``MAE" and the ``$\text{Error}_{1.5}$" lines for ``CoSaMP" in Figure~\ref{fig:PlotW} appear beyond the scope of the vertical axis. It can be seen that our metric-optimal algorithm outperforms the two other methods.

\begin{figure}[t]
\centering
\begin{subfigure}[t]{1.0\linewidth}
\centering
\includegraphics[width=8.6cm]{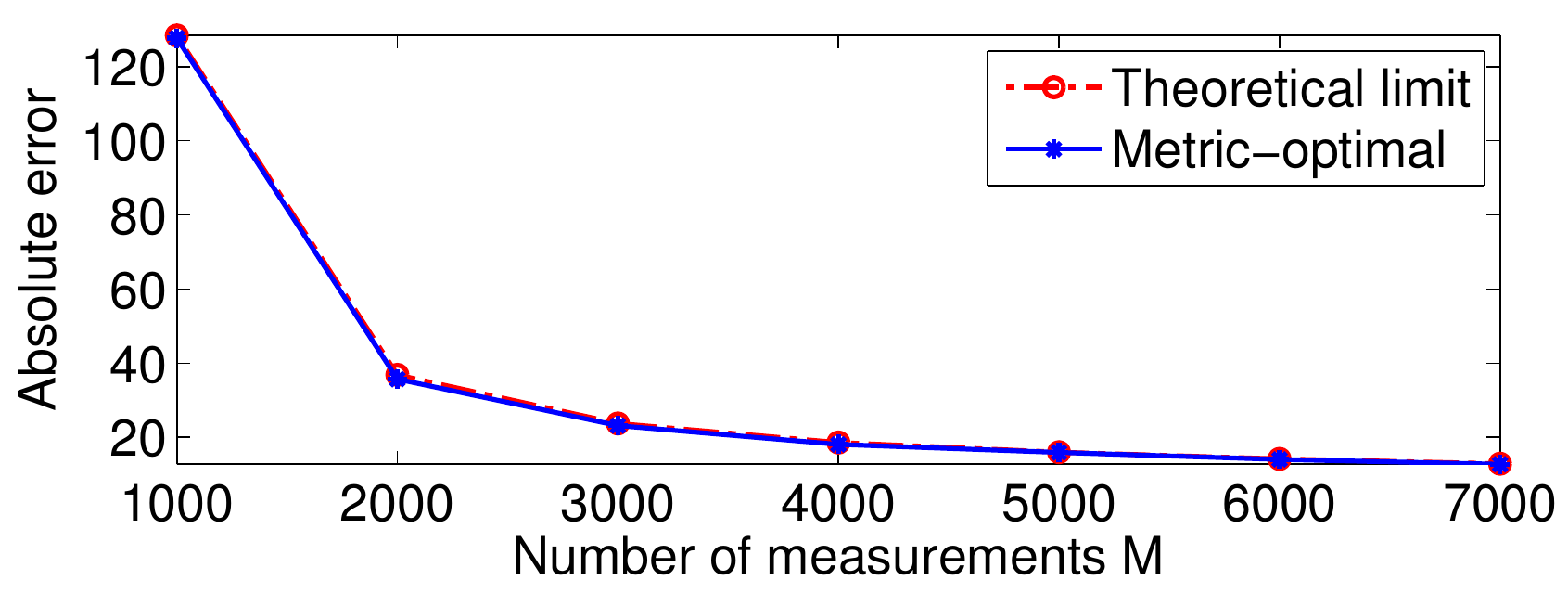}
\caption{\footnotesize\sl Absolute error.}
\label{fig:plot_ell1}
\end{subfigure}
\begin{subfigure}[t]{1.0\linewidth}
\centering
\includegraphics[width=8.6cm]{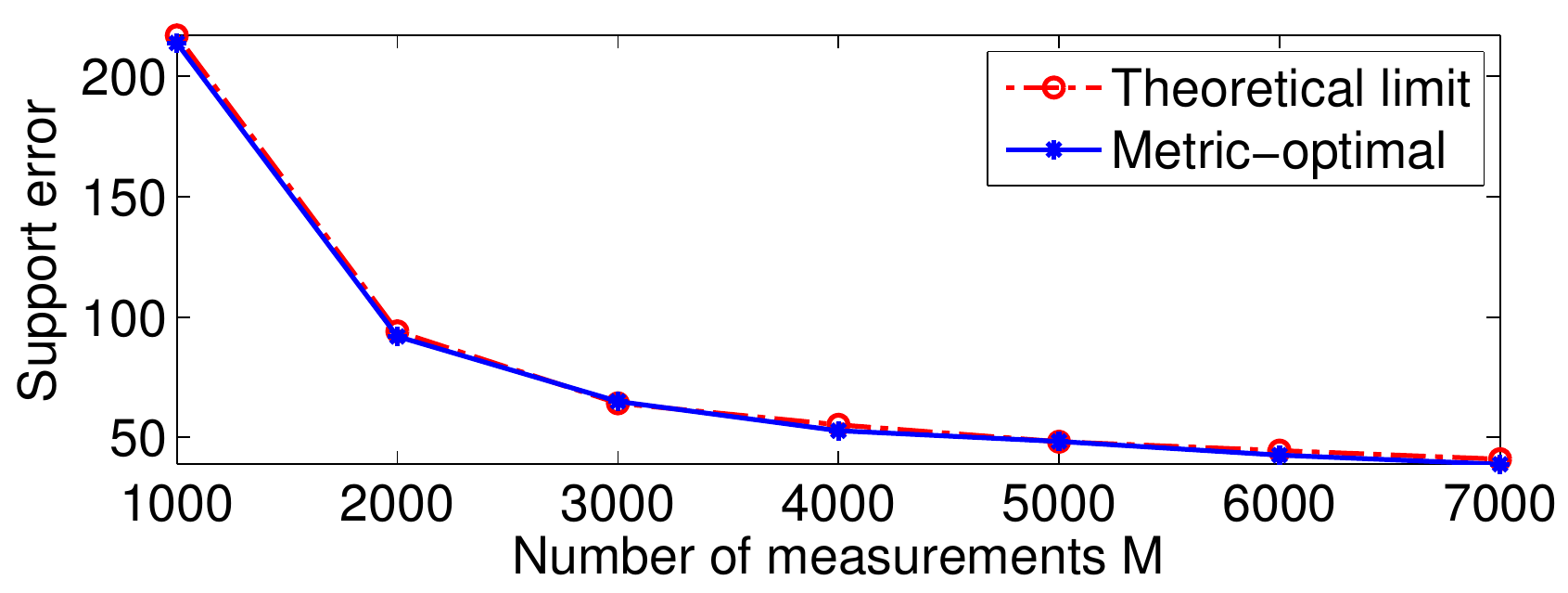}
\caption{\footnotesize\sl Support error.}
\label{fig:plot_ell0}
\end{subfigure}
\caption{
{\small\sl
Comparisons of the metric-optimal estimators and the corresponding theoretical limits~\eqref{eq:MMAEBound}, and~\eqref{eq:supportBound}. The corresponding lines coincide. (Sparse Gaussian input and Gaussian channel; sparsity rate~$s=3\%$; input length N=10,000; SNR=20dB.)
}
\label{fig:Plot_theo}
}
\end{figure}

To demonstrate the theoretical analysis of our algorithm in Section~\ref{subsec:Eg}, we compare our MMAE estimation results with the theoretical limit~\eqref{eq:MMAEBound} in Figure~\ref{fig:plot_ell1}, where the integrations are computed numerically. In Figure~\ref{fig:plot_ell0}, we compare our MMSuE estimator with the theoretical limit~\eqref{eq:supportBound}, where the value of~$\mu$ is acquired numerically from the relaxed BP method~\cite{Rangan:web:GAMP} with 20 iterations.
In these two figures, each point on the ``metric-optimal" line is generated by averaging 40 experiments with the same parameters. It is shown in both figures that the lines coincide. Therefore our estimation algorithm reaches the corresponding theoretical limits and is optimal.

\subsection{The $\ell_\infty$ norm error}
\label{eq:infty_Sim}
\begin{figure}[t]
\centering
\includegraphics[width=8cm]{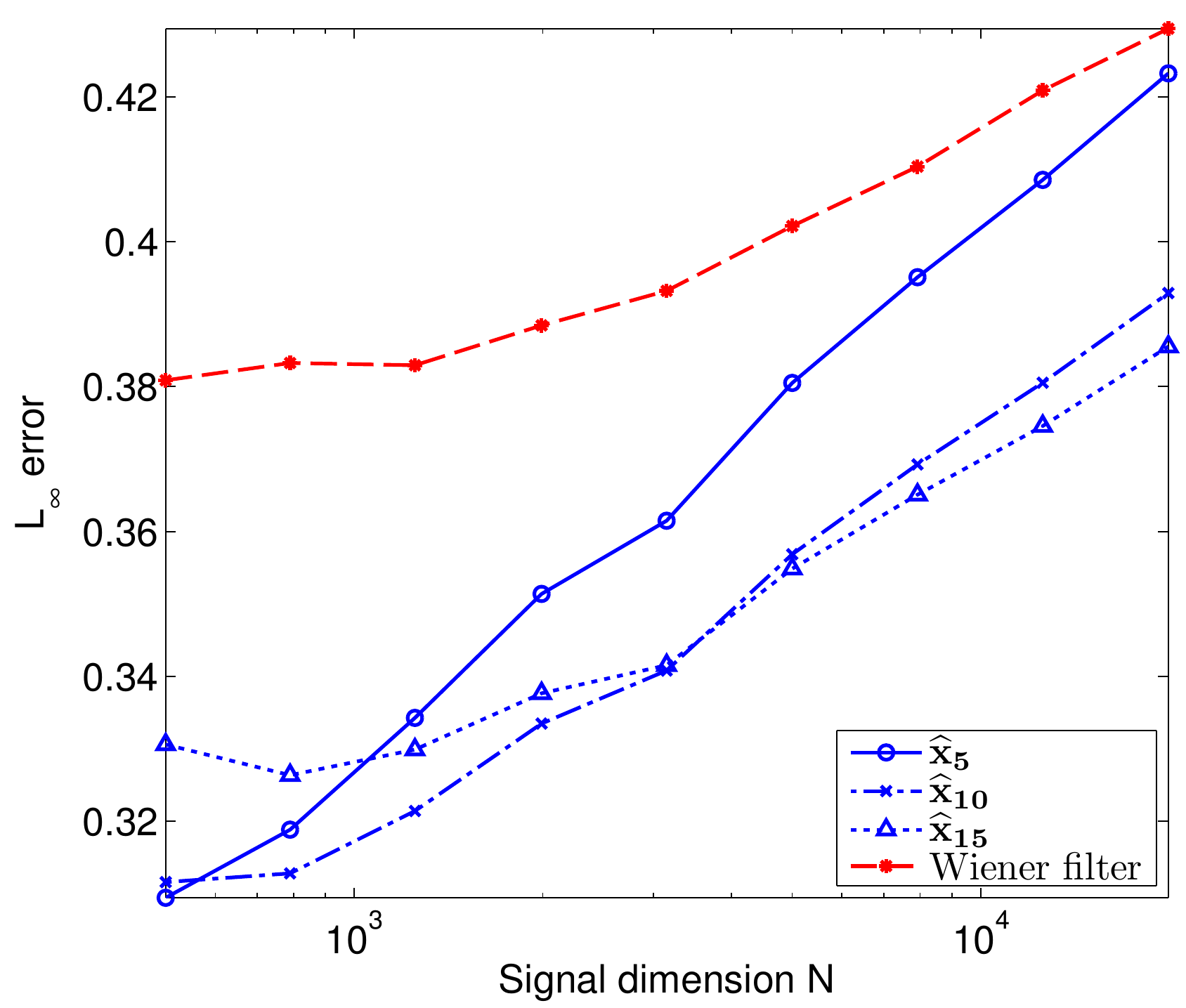}
\caption{
{\small\sl
The performance of the Wiener filter as well as the estimators~$\mathbf{\widehat{x}}_5$, $\mathbf{\widehat{x}}_{10}$, and $\mathbf{\widehat{x}}_{15}$ in terms of $\ell_\infty$ error. (Sparse Gaussian input and Gaussian channel; sparsity rate~$s=5\%$; measurement ratio $M/N=0.3$; SNR=20dB.)
}
\label{fig:N_VS_L_infty}
}
\end{figure}

To show the performance of the Wiener filter and our metric-optimal algorithm in terms of the $\ell_\infty$ error,  we test the Wiener filter~\eqref{eq:c_q} as well as the estimator~\eqref{eq:xhat_p}. The simulation settings remain the same with the previous simulations, except that we set the sparsity rate $s=5\%$, and we only test the sparse Gaussian signal. The ratio $M/N$ is fixed to $0.3$, and the signal dimension $N$ ranges from $500$ to $20,000$. We then run our metric-optimal algorithm with $p=5, 10, 15$ in \eqref{eq:xhat_p}. In Figure~\ref{fig:N_VS_L_infty}, we present  the $\ell_\infty$ norm error of the Wiener filter~\eqref{eq:c_q}, along with the $\ell_\infty$ norm error of $\xhat_{5}$, $\xhat_{10}$, and $\xhat_{15}$~\eqref{eq:xhat_p}. It can be seen that the estimators~$\xhat_{5}$, $\xhat_{10}$, and $\xhat_{15}$ all achieve lower $\ell_\infty$ errors than the Wiener filter~\eqref{eq:min_infty} does, because the signal dimension is finite. Thus, for a finite signal dimension, it is reasonable to choose a proper value of $p$, and run the metric-optimal algorithm for $\mathbf{\widehat{x}}_p$~\eqref{eq:xhat_p} to achieve a low $\ell_\infty$ error. That said, these results are specific to sparse Gaussian inputs, and it remains to be seen whether they will carry over to other input distributions.

\section*{Acknowledgment}
We thank Sundeep Rangan for kindly providing the Matlab code \cite{Rangan:web:GAMP} of the relaxed BP algorithm \cite{Rangan2010,RanganGAMP2010}.

{\small
\bibliographystyle{IEEEbib}
\bibliography{cites}

\begin{thebibliography}{10}

\bibitem{Tan2012SSP}
J.~Tan, D.~Carmon, and D.~Baron,
\newblock ``Optimal estimation with arbitrary error metric in compressed
  sensing,''
\newblock in {\em Proc. IEEE Stat. Sig. Proc. Workshop (SSP)}, Aug. 2012, pp.
  588--591.

\bibitem{Tan2012signal}
J.~Tan, D.~Carmon, and D.~Baron,
\newblock ``Signal estimation with arbitrary error metrics in compressed
  sensing,''
\newblock {\em arXiv:1207.1760}, July 2012.

\bibitem{Tan2013}
J.~Tan, D.~Baron, and L.~Dai,
\newblock ``Minimax error estimation in linear mixing sytems,''
\newblock in preparation.

\bibitem{CandesRUP}
E.~Cand\`{e}s, J.~Romberg, and T.~Tao,
\newblock ``Robust uncertainty principles: {E}xact signal reconstruction from
  highly incomplete frequency information,''
\newblock {\em IEEE Trans. Inf. Theory}, vol. 52, no. 2, pp. 489--509, Feb.
  2006.

\bibitem{DonohoCS}
D.~Donoho,
\newblock ``Compressed sensing,''
\newblock {\em IEEE Trans. Inf. Theory}, vol. 52, no. 4, pp. 1289--1306, Apr.
  2006.

\bibitem{Huber1973}
P.J. Huber,
\newblock ``{Robust regression: asymptotics, conjectures and Monte Carlo},''
\newblock {\em Ann. Stat.}, vol. 1, no. 5, pp. 799--821, 1973.

\bibitem{OLeary1990}
D.P. O'Leary,
\newblock ``Robust regression computation using iteratively reweighted least
  squares,''
\newblock {\em SIAM J. Matrix Anal. Appl.}, vol. 11, no. 3, pp. 466--480, July
  1990.

\bibitem{GuoWang2008}
D.~Guo and C.-C. Wang,
\newblock ``Multiuser detection of sparsely spread {CDMA},''
\newblock {\em IEEE J. Sel. Areas Commun.}, vol. 26, no. 3, pp. 421--431, Apr.
  2008.

\bibitem{GuoWang2007}
D.~Guo and C.C. Wang,
\newblock ``Random sparse linear systems observed via arbitrary channels: A
  decoupling principle,''
\newblock in {\em Proc. Int. Symp. Inf. Theory (ISIT2007)}, June 2007, pp.
  946--950.

\bibitem{Rangan2010}
S.~Rangan,
\newblock ``Estimation with random linear mixing, belief propagation and
  compressed sensing,''
\newblock {\em CoRR}, vol. arXiv:1001.2228v1, Jan. 2010.

\bibitem{Grenander1957}
U.~Grenander and M.~Rosenblatt,
\newblock {\em Statistical analysis of stationary time series},
\newblock Wiley New York, 1957.

\bibitem{Levy2008}
B.C. Levy,
\newblock {\em Principles of signal detection and parameter estimation},
\newblock Springer Verlag, 2008.

\bibitem{Cover91}
T.~M. Cover and J.~A. Thomas,
\newblock {\em Elements of Information Theory},
\newblock Wiley-Interscience, 1991.

\bibitem{Webb2002}
A.R. Webb,
\newblock {\em Statistical pattern recognition},
\newblock John Wiley \& Sons Inc., 2002.

\bibitem{GuoVerdu2005}
D.~Guo and S.~Verd{\'u},
\newblock ``Randomly spread {CDMA}: {A}symptotics via statistical physics,''
\newblock {\em IEEE Trans. Inf. Theory}, vol. 51, no. 6, pp. 1983--2010, June
  2005.

\bibitem{Guo2006}
D.~Guo and C.C. Wang,
\newblock ``Asymptotic mean-square optimality of belief propagation for sparse
  linear systems,''
\newblock in {\em IEEE Inf. Theory Workshop}, Oct. 2006, pp. 194--198.

\bibitem{GuoBaronShamai2009}
D.~Guo, D.~Baron, and S.~Shamai,
\newblock ``A single-letter characterization of optimal noisy compressed
  sensing,''
\newblock in {\em Proc. 47th Allerton Conf. Commun., Control, and Comput.},
  Sep. 2009.

\bibitem{RFG2009}
S.~Rangan, A.~K. Fletcher, and V.~K. Goyal,
\newblock ``Asymptotic analysis of {MAP} estimation via the replica method and
  applications to compressed sensing,''
\newblock {\em CoRR}, vol. abs/0906.3234, June 2009.

\bibitem{CSBP2010}
D.~Baron, S.~Sarvotham, and R.~G. Baraniuk,
\newblock ``Bayesian compressive sensing via belief propagation,''
\newblock {\em IEEE Trans. Signal Process.}, vol. 58, pp. 269--280, Jan. 2010.

\bibitem{RanganGAMP2010}
S.~Rangan,
\newblock ``Generalized approximate message passing for estimation with random
  linear mixing,''
\newblock {\em Arxiv preprint arXiv:1010.5141}, Oct. 2010.

\bibitem{TroppOMP}
J.~A. Tropp and A.~C. Gilbert,
\newblock ``Signal recovery from random measurements via orthogonal matching
  pursuit,''
\newblock {\em IEEE Trans. Inf. Theory}, vol. 53, no. 12, pp. 4655--4666, Dec.
  2007.

\bibitem{Cosamp08}
D.~Needell and J.~A. Tropp,
\newblock ``Co{S}a{MP}: Iterative signal recovery from incomplete and
  inaccurate samples,''
\newblock {\em Appl. Comput. Harm. Anal.}, vol. 26, no. 3, pp. 301--321, 2008.

\bibitem{indyk2008near}
P.~Indyk and M.~Ruzic,
\newblock ``Near-optimal sparse recovery in the $\ell_1$ norm,''
\newblock in {\em 49th Annu. IEEE Symp. Found. Comput. Sci.}, Oct. 2008, pp.
  199--207.

\bibitem{berinde2008practical}
R.~Berinde, P.~Indyk, and M.~Ruzic,
\newblock ``Practical near-optimal sparse recovery in the $\ell_1$ norm,''
\newblock in {\em Proc. 46th Allerton Conf. Commun., Control, and Comput.},
  Sep. 2008, pp. 198--205.

\bibitem{CDDNOA}
A.~Cohen, W.~Dahmen, and R.~A. DeVore,
\newblock ``Near optimal approximation of arbitrary vectors from highly
  incomplete measurements,''
\newblock 2007,
\newblock preprint.

\bibitem{Wang2010}
W.~Wang, M.J. Wainwright, and K.~Ramchandran,
\newblock ``Information-theoretic limits on sparse signal recovery: Dense
  versus sparse measurement matrices,''
\newblock {\em IEEE Trans. Inf. Theory}, vol. 56, no. 6, pp. 2967--2979, June
  2010.

\bibitem{Tulino2011}
A.~Tulino, G.~Caire, S.~Shamai, and S.~Verd{\'u},
\newblock ``Support recovery with sparsely sampled free random matrices,''
\newblock in {\em IEEE Int. Symp. Inf. Theory}, 2011, pp. 2328--2332.

\bibitem{Wainwright2009}
M.J. Wainwright,
\newblock ``Information-theoretic limits on sparsity recovery in the
  high-dimensional and noisy setting,''
\newblock {\em IEEE Trans. Inf. Theory}, vol. 55, no. 12, pp. 5728--5741, Dec.
  2009.

\bibitem{Akcakaya2010}
M.~Ak{\c{c}}akaya and V.~Tarokh,
\newblock ``Shannon-theoretic limits on noisy compressive sampling,''
\newblock {\em IEEE Trans. Inf. Theory}, vol. 56, no. 1, pp. 492--504, Jan.
  2010.

\bibitem{Reeves2011sampling}
G.~Reeves and M.~Gastpar,
\newblock ``The sampling rate-distortion tradeoff for sparsity pattern recovery
  in compressed sensing,''
\newblock {\em IEEE Trans. Inform. Theory}, vol. 58, no. 5, pp. 3065--3092, May
  2012.

\bibitem{cadzow2002}
J.A. Cadzow,
\newblock ``Minimum $\ell_1$, $\ell_2$, $\ell_\infty$ norm approximate
  solutions to an overdetermined system of linear equations,''
\newblock {\em Digital Signal Proces.}, vol. 12, no. 4, pp. 524--560, Oct.
  2002.

\bibitem{clark1961}
C.E. Clark,
\newblock ``The greatest of a finite set of random variables,''
\newblock {\em Oper. Res.}, vol. 9, no. 2, pp. 145--162, Mar. 1961.

\bibitem{indyk2001}
P.~Indyk,
\newblock ``On approximate nearest neighbors under $\ell_\infty$ norm,''
\newblock {\em J. Comput. Syst. Sci.}, vol. 63, no. 4, pp. 627--638, Dec. 2001.

\bibitem{Lounici2008}
K.~Lounici,
\newblock ``Sup-norm convergence rate and sign concentration property of lasso
  and {D}antzig estimators,''
\newblock {\em Electron. J. Stat.}, vol. 2, pp. 90--102, 2008.

\bibitem{Bishop2006}
C.M. Bishop,
\newblock {\em Pattern recognition and machine learning}, vol.~4,
\newblock Springer New York, 2006.

\bibitem{Caire2004}
G.~Caire, R.R. Muller, and T.~Tanaka,
\newblock ``Iterative multiuser joint decoding: Optimal power allocation and
  low-complexity implementation,''
\newblock {\em Information Theory, IEEE Transactions on}, vol. 50, no. 9, pp.
  1950--1973, Sep. 2004.

\bibitem{Montanari2006}
A.~Montanari and D.~Tse,
\newblock ``Analysis of belief propagation for non-linear problems: The example
  of {CDMA} (or: How to prove {T}anaka's formula),''
\newblock in {\em IEEE Inf. Theory Workshop}, Mar. 2006, pp. 160--164.

\bibitem{Rangan2011}
S.~Rangan, A.K. Fletcher, V.K. Goyal, and P.~Schniter,
\newblock ``Hybrid approximate message passing with applications to structured
  sparsity,''
\newblock {\em Arxiv preprint arXiv:1111.2581}, Nov. 2011.

\bibitem{Tanaka2002}
T.~Tanaka,
\newblock ``{A statistical-mechanics approach to large-system analysis of CDMA
  multiuser detectors},''
\newblock {\em IEEE Trans. Inf. Theory}, vol. 48, no. 11, pp. 2888--2910, Nov.
  2002.

\bibitem{DMM2009}
D.~L. Donoho, A.~Maleki, and A.~Montanari,
\newblock ``{Message passing algorithms for compressed sensing},''
\newblock {\em Proc. Nat. Acad. Sci.}, vol. 106, no. 45, pp. 18914--18919, Nov.
  2009.

\bibitem{Rangan:web:GAMP}
S.~Rangan, A.~Fletcher, V.~Goyal, U.~Kamilov, J.~Parker, and P.~Schniter,
\newblock ``{GAMP},''
  \url{http://gampmatlab.wikia.com/wiki/Generalized_Approximate_Message_Passing/}.

\bibitem{Wiener1949}
N.~Wiener,
\newblock {\em Extrapolation, interpolation, and smoothing of stationary time
  series with engineering applications},
\newblock MIT press, 1949.

\end{thebibliography}
}
\end{document}